\documentclass[a4paper,fleqn,usenatbib]{mnras}
\usepackage[T1]{fontenc}
\usepackage{ae,aecompl}
\usepackage{graphicx}
\usepackage{footnote}
\usepackage[fleqn]{amsmath}
\usepackage[varg]{txfonts}
\usepackage{multirow}
\usepackage{subfig}
\newcommand\numberthis{\addtocounter{equation}{1}\tag{\theequation}}

\newcommand{\aec}{A\&C}

\title[Calculating transient rates from surveys]{Calculating transient rates from surveys}
\author[D.~Carbone et al.]{
D.~Carbone$^{1}$\thanks{E-mail: d.carbone@uva.nl}, A.~J.~van~der~Horst$^{2}$, R.~A.~M.~J.~Wijers$^{1}$, A.~Rowlinson$^{1,3}$ \\
$^{1}$ Anton Pannekoek Institute for Astronomy, University of Amsterdam, Postbus 94249, 1090 GE Amsterdam, The Netherlands\\
$^{2}$ Department of Physics, The George Washington University, 725 21$^{st}$ Street NW, Washington, DC 20052, USA\\
$^{3}$ ASTRON, The Netherlands Institute for Radio Astronomy, Postbus 2, 7990 AA Dwingeloo, The Netherlands
}

\date{Accepted XXX. Received YYY; in original form ZZZ}
\pubyear{2016}

\begin{document}
\label{firstpage}
\pagerange{\pageref{firstpage}--\pageref{lastpage}}
\maketitle

\begin{abstract}
We have developed a method to determine the transient surface density and transient rate for any given survey, using Monte-Carlo
simulations. 
This method allows us to determine the transient rate as a function of both the flux and the duration of the transients in the whole
flux-duration plane rather than one or a few points as currently available methods do.
It is applicable to every survey strategy that is monitoring the same part of the sky, 
regardless the instrument or wavelength of the survey, or the target sources.
We have simulated both top-hat and Fast Rise Exponential Decay light curves, highlighting how the shape of the light curve might
affect the detectability of transients. 
Another application for this method is to estimate the number of transients of a given kind that are expected to be detected by a survey,
provided that their rate is known.
\end{abstract}


\section{Introduction}
\label{sec:intro}

Sources of transient emission are observed over many timescales, from years down to a fraction of a second, and in every band of the
electromagnetic spectrum, from high-energy $\gamma$-rays to low-frequency radio waves. Studying this emission is very important to
understand the dynamics of the Universe and its constituents, and can reveal the physics behind some of the most extreme
phenomena and objects. 
One of the most basic quantities for understanding the nature and physics of such sources is the number of transients per unit area,
i.e. the transient surface density. 
When one also invokes a certain timescale of occurrence, this can be converted into a transient rate.
While the transient surface density is straightforward to calculate, since the total area of a given survey is typically well known,
determining the transient rate is not trivial because it involves both the timescales of the survey and the duration of the transients
\citep[see, e.g.,][]{PanSTARRS_Carbone2014}

The transient surface density, as well as the transient rate, can be calculated as a function of various parameter of the transients,
e.g., their flux and duration. 
If a survey results in non-detections, the common way to calculate upper limits for the transient surface density as a function of flux
is to assume a Poisson distribution for transients, and use all the independent pairs of observations within the survey to calculate
the total amount of surveyed area. 
This method allows us to set an upper limit to the transient surface density of transients that could have been detected in every image
in the dataset, therefore the flux limit is set by the sensitivity of the noisiest image
\citep[see, e.g.,][]{VLAarchival_Bell2011, SN&trans_Alexander2014}.

\citet{PanSTARRS_Carbone2014} developed a method to calculate the transient rate as a function of the duration of the transients.
They calculated the number of statistically independent pairs of observations for each duration of the transients they considered,
governed by the setup of their survey, and determined the upper limits on the transient surface density for different durations.
In this process they also applied a correction to take into account the probability that transients could fall in gaps between observations.
While this was already an improvement upon other methods in the literature, it was still limited in the sense that when calculating
the transient surface density as a function of flux or as a function of duration, only one of the two variables was taken into account
and the other one ignored. 
For instance, regarding the aforementioned correction on transients falling in gaps between observations,
\citet{PanSTARRS_Carbone2014} assumed that if a transient were on during an observation, it would be always detected. 
This is not necessarily true, because it could have been detectable only during a fraction of the observation, and therefore its flux
would have been averaged over the whole duration of the observation and its signal smeared out so much that it could not be
detected in the end.
In this paper we are taking the next step in accurately determining the transient rate as a function of flux and duration, for any
specified survey, by performing Monte-Carlo simulations with transients of varying flux and duration, and also for different light
curve shapes.

Simulation work has been done by other groups, but their scopes and aims were different than the work presented in this paper. 
\citet{Rates_Trott2013} developed a framework to calculate the transient rate in beam-formed data (i.e., not in images), starting
from technical parameters of the survey and assuming a flux distribution for the astrophysical source population. 
Their detailed framework can take into account many parameters of a radio observation (e.g., bandwidth, beam shape, system
temperature) in order to calculate the minimum flux a transient must have to be detectable, and convert this into a rate of events
for fast radio transient sources. 
Image plane transients and transients at other wavelengths can not be well represented by these simulations.
\citet[][]{SKA_memo_Cordes2007} developed a method to determine the number of transient sources of a specified population
that should be detected in a given survey, starting from the properties of the transient population as well as the properties of
the survey. 
The relevant source properties in their simulations are luminosity, duration of the transient, period, rate, and number density;
the properties of the observing campaign are sampled area, duration and speed of the observations, and characteristics of
the instrument (such as noise and resolution). 
From these parameters it is possible to estimate the probability that a transient source is detectable when it is observed, and
the total number of sources that should be detected. 
While their approach is aimed at transient searches in images, it is targeted to a specific transient population and does not
take into account the fact that a transient survey will detect transients of
the combination of all populations of transient sources.

In this paper we present a new method to perform Monte-Carlo simulations of transient detection, with no constraints on the
observing strategy nor the observing frequency. 
This method applies to surveys at all the wavelengths and transient sources of every type.
We describe the method in Section~\ref{sec:method} and the setup of the simulations in Section~\ref{sec:setup}.
In Section~\ref{sec:noerrors} we show our results in an ideal case, to validate our method, and in Section~\ref{sec:real}
we present a realistic situation. 
We discuss our findings, comparing them to previous methods, in Section~\ref{sec:discussion}, and summarise our results
and discussion in Section~\ref{sec:conclusion}.

\section{Description of the method}
\label{sec:method}

In our simulations, a source is fully characterised by its peak flux and duration.
This is because our method assumes that all the images have the same field of view and pointing centre, and that the noise in the
images is uniform.
Both flux and duration are uniformly distributed in log space between user-specified boundaries. A spread in the flux is also
simulated in order to take into account measurement errors; this implies that the flux of every source is different in every
observation.
The start time of each transient is simulated uniformly distributed between the beginning of the survey minus its
duration and the end of the survey.
By definition, the survey starts at the start of the first observation and ends at the end of the last one.
The duration of the survey (T$_{survey}$) is defined as the difference between these two times.

The aim of our simulations is to check if transients are detected as such in a given survey.
To do so, each simulated transient is checked against every observation to verify if it was on during that time interval.
If it was, the flux of the transient within the observation is determined and compared to the sensitivity of the observation.
If it is higher, this means that the transient is detected in that observation.

In real transient surveys we want to distinguish if a source that is newly detected is a real transient or a fake source
(for example a noise peak), or a constant source that was previously missed.
In order to do so, we implemented a technique that is the same as currently used by the LOFAR
\textsc{Transient Pipeline}
\citep{trap_Swinbank2015}. This technique consists of checking if the new source would have been detected in the
best image in the dataset before it is detected, using a higher threshold to make sure that it is not a noise fluctuation.
In this way it is established if that source should have been detected earlier or not.
After all observations have been checked, sources that are never detected or could have been missed are flagged as non detected;
sources that have been always detected are flagged as constant; sources that could not be missed in some observations but have not
been detected in others are flagged as transients.

Once we establish which transients have been detected we can perform statistical calculations.
A logarithmic grid is calculated both in flux and duration, and populated with sources, keeping track of those that were flagged
as transients.
For each combination of flux and duration we calculate the probability of detecting a transient as a simple ratio between the
number of transients detected and the total number of sources for that particular combination (N$_{\textrm{det}}$/N$_{\textrm{all}}$).

Finally, we translate the probability into a transient rate.
To do so, we calculate the transient rate we simulated ($\hat{\rho}_{\textrm{sim}}^{\textrm{all}}$) for each combination of flux F and
duration T:
\begin{equation}
\hat{\rho}_{\textrm{sim}}^{\textrm{all}} (F, T) = \frac{N_{\textrm{sim}}^{\textrm{all}} (F, T)}{(T_{\textrm{survey}} + T) \times FOV},
\end{equation}

\noindent where FOV is the field of view
of each observation. Again, the assumption here is that we are always observing the same field.

Since we do not detect all the transients we simulated, the rate of transients detected ($\hat{\rho}_{\textrm{sim}}^{\textrm{det}}$)
can be expressed as:
\begin{equation}
\hat{\rho}_{\textrm{sim}}^{\textrm{det}} (F, T) = \frac{N_{\textrm{sim}}^{\textrm{det}} (F, T)}{(T_{\textrm{survey}} + T) \times FOV} .
\end{equation}

We also define the probability ($p$) of detecting a transient as:
\begin{align*}
p (F, T) &= \frac{N_{\textrm{sim}}^{\textrm{det}} (F, T)}{N_{\textrm{sim}}^{\textrm{all}} (F, T)} =
\frac{N_{\textrm{real}}^{\textrm{det}} (F, T)}{N_{\textrm{real}}^{\textrm{all}} (F, T)} \\
&= \frac{\hat{\rho}_{\textrm{sim}}^{\textrm{det}} (F, T)}{\hat{\rho}_{\textrm{sim}}^{\textrm{all}} (F, T)} = 
\frac{\hat{\rho}_{\textrm{real}}^{\textrm{det}} (F, T)}{\hat{\rho}_{\textrm{real}}^{\textrm{all}} (F, T)}  . \numberthis
\label{eq:probability}
\end{align*}

After performing a survey, a certain number of transients with flux F and duration T will be detected
($N_{\textrm{real}}^{\textrm{det}}$), and using Equation~\ref{eq:probability} we can estimate how
many transients occurred in total during the survey ($N_{\textrm{real}}^{\textrm{all}}$).
Then we can estimate the real rate of transients ($\rho_{\textrm{real}}^{\textrm{all}}$) using the relation:
\begin{align*}
\hat{\rho}_{\textrm{real}}^{\textrm{all}} (F, T) & = \hat{\rho}_{\textrm{sim}}^{\textrm{all}} (F, T) \times
\frac{N_{\textrm{real}}^{\textrm{all}} (F, T)}{N_{\textrm{sim}}^{\textrm{all}} (F, T)} \\
& = \frac{N_{\textrm{real}}^{\textrm{all}} (F, T)}{{(T_{\textrm{survey}} + T) \times FOV}} \\
& = \frac{N_{\textrm{real}}^{\textrm{det}} (F, T)}{{p (F, T) \times (T_{\textrm{survey}} + T) \times FOV}} . \numberthis
\label{eq:tr_rate}
\end{align*}

If no transients have been detected in our survey, we can derive an upper limit on the transient rate.
To calculate the 95\% confidence level upper limit on the transient rate from our survey, we
assume a Poisson distribution:
\begin{equation}
P (n; \lambda) = e^{-n} \frac{n^{\lambda}}{n!} , 
\label{eq:poisson}
\end{equation}

\noindent where $\lambda$ is the expectation value and n is the number of realisations. In this case n\,=\,0 and $\lambda$
is the maximum number of transient that could have been detected ($N^{\textrm{det, ul}}_{\textrm{real}}$).
The 95\% confidence level is defined as P(0; $\lambda$)\,=\,0.05, and thus:
\begin{align*}
P (0; N^{\textrm{det, ul}}_{\textrm{real}}(F, T)) & = e^{-N^{\textrm{det, ul}}_{\textrm{real}}(F, T)} = 0.05 , \\
N^{\textrm{det, ul}}_{\textrm{real}} (F, T) & = \textrm{ln}(20) = 3.00 . \numberthis
\end{align*}

We can now insert this into Equation~\ref{eq:tr_rate} to obtain the upper limit on the transient rate with confidence level \textit{c}:

\begin{align*}
\hat{\rho}_{\textrm{real}}^{\textrm{all, ul}} (F, T) & = \frac{N_{\textrm{real}}^{\textrm{det, ul}} (F, T)}{{p (F, T)
\times (T_{\textrm{survey}} + T) \times FOV}} \\
& = -\frac{\textrm{ln}(1\,-\,c)}{{p (F, T) \times (T_{\textrm{survey}} + T) \times FOV}} . \numberthis
\label{eq:tr_rate_ul}
\end{align*}

To summarise, if there are detections of transients in a given survey, the transient rate can be determined by using
Equation~\ref{eq:tr_rate}; in the case of non-detection the upper limit with confidence level \textit{c} can be estimated with
Equation~\ref{eq:tr_rate_ul}.
The transient surface density can be calculated using the same equations multiplied by
$(T_{survey} + T)$.

\section{Setup of the simulations}
\label{sec:setup}

In order to run our simulations, we need to have an observing setup.
The information required is the start time, the duration and the sensitivity of every image.
The images have to cover the same field of view as it is assumed that every transient is observable from every image.
Other input that is required is the number of transients to be simulated, minimum and maximum flux and duration,
and fractional errors on the flux.
Two more parameters which have to be defined are the detection threshold and the extra threshold.
The detection threshold defines the minimum signal to noise ratio that a source has to have in order to be detected.
Because of the fact that different images have different sensitivities, a constant source just brighter than the threshold
can be detectable in some images and missed in others, and therefore considered to be a transient.
This could lead to many false transient detections.
To avoid this, a source is defined transient only if it would have been detected in the 
best image in the dataset before the source is detected, using a higher extra threshold.

In our simulations we simulated transients with two different light curve shapes: top-hat and Fast Rise Exponential Decay (FRED).
The top-hat transients have a flux that jumps from zero to a certain value instantaneously, stays constant for the duration of the
transient, and drops to zero instantaneously.
FREDs have a flux that rises from zero to a maximum flux instantaneously, but the decay of their flux is exponential and it is
determined by its e-folding time, i.e., F(t) = F$_0\,\times\,e^{-t\,/\,\tau}$.
In our description we define the duration of a FRED as its e-folding time.

\subsection{Survey strategy}
\label{sec:survey_lofar}
Our simulations are set up in such a way that any survey strategy, abiding the aforementioned requirement, can be used.
In this paper we chose to use the observing setup described in \citet{PanSTARRS_Carbone2014}.
These observations were taken using the LOw Frequency ARray \citep[LOFAR;][]{LOFAR_vanHaarlem2013} between March
and August 2013. In this observing campaign four different fields were monitored but for this work only one of those is used.
Each observation covers a field of view of 15.48\,deg$^2$ and has a duration of 11\,min.
Some of the images have been discarded because they were corrupted by Radio Frequency Interference.
The exact list of observing dates is reported in Table~2 of \citet{PanSTARRS_Carbone2014} (target MD03).
The sensitivity of the images ranges from 0.21\,Jy\,beam$^{-1}$ to 0.42\,Jy\,beam$^{-1}$.
We have simulated 2 million sources with fluxes between 0.1 and 100\,Jy, and durations between 30\,sec and 2000\,days.
A detection threshold of 8 and an extra threshold of 3 are used, meaning that sources brighter than 8 times the noise are
detected, while only the ones above 8+3 times the noise of the best previous image are considered real detections.

\subsection{Known regimes}
\label{sec:known}
In the flux-duration plane we can give predictions on the value of the probability of detecting transients or of its trend in specific regimes.
In order to give explicit expressions for each of these regimes, we first define some parameters.
S$_{max}$ is the sensitivity of the noisiest image in the dataset, S$_{min}$ is the sensitivity of the best image in the dataset,
S$_{extra}$ is the sensitivity of the best image including the extra threshold,
T$_{max\_gap}$ is the maximum gap between two consecutive observations, T$_{survey}$ is the total duration
of the observing campaign, T$_{on}$ is the total time spent observing during the survey, finally, S$_{obs}$ and T$_{obs}$ are the sensitivity
and integration time of a single observation.
Here we discuss the different regimes:
\begin{itemize}
\item Sources with a flux smaller than the sensitivity of the best observation are never detectable: S\,$<$\,S$_{min}$, p(F, T)\,=\,0.
\item Sources longer than the maximum separation between two consecutive observations but shorter than the total duration of the
survey, and brighter than the sensitivity of the best observation including the extra threshold are always detected as transients:
T$_{max\_gap}$\,$<$\,T\,$<$\,T$_{survey}$ and S\,$>$\,S$_{extra}$, p(F, T)\,=\,1.
The boundary in flux of this region does not actually coincide exactly with the sensitivity of the best image in the dataset, including the extra
threshold, but it lies between this value and S$_{max}$.
The exact value depends on the survey setup, i.e., when in the dataset the best image is: if the best image is the very first, than the
boundary is sharply at its sensitivity corrected for the extra threshold, otherwise it is above it.
\item Sources longer than the total duration of the survey and brighter than the sensitivity of the best observation
can be considered constant sources. In these simulations a source can start between one duration before the beginning of the survey
starts and when the survey is over; this implies that the time interval when sources can start is T\,+\,T$_{survey}$. If the duration of
the transient is longer than the duration of the survey there is an interval of length T\,--\,T$_{survey}$ when the transient starts before
the survey and finishes after that is over. If the transient starts in this period it is treated as a constant source, if it start elsewhere it is
detected as a transient.
The interval when it is recognised as transient can be expressed as (T\,+\,T$_{survey}$)\,--\,(T\,--\,T$_{survey}$)\,=\,2\,T$_{survey}$.
The probability of detecting a source as a transient is therefore the ratio between the interval when it is detected as such and the total
interval when the source can start.
T\,$>$\,T$_{survey}$ and S\,$>$\,S$_{max}$, p(F, T)\,=\,$\frac{2\,T_{survey}}{T\,+\,T_{survey}}$.
\item The probability to detect sources much shorter than the duration of one observation with a flux so high that if they are in
an observation they are detectable is equal to the total amount of time spent observing with respect to the total duration of the survey:
T\, $<<$\, T$_{obs}$
and S\,$>>$\,S$_{max}$, p(F, T)\,=\,$\frac{T_{on}}{T_{survey}}$.
\item Sources shorter than the duration of an observation are detectable only if their flux is high enough so that their fluence is
higher than the minimum required. The minimum fluence is set by the product of the sensitivity of the observation and its duration;
assuming a top-hat lightcurve, the flux a transient falling completely in an observation should have in order to be detected is:
S\,$\ge$\,S$_{obs} \frac{T_{obs}}{T}$.
\end{itemize}

\section{Ideal case}
\label{sec:noerrors}
In order to validate our method and code, we ran our simulations without any error in the flux measurements;
this means that every observation measures the correct flux for every source, without any scatter.
To demonstrate that our code works correctly, we are showing here the results of our simulations with top-hat transients.
We note that we also verified the simulations with FREDs, but we do not show the resulting plots due to redundancy.

Using the simulations setup described in Section~\ref{sec:setup} and the simulations method described in
Section~\ref{sec:method}, we have calculated the probability for detecting transients, transient surface density and transient rate,
as a function of flux and duration.
The top panel of Figure~\ref{fig:noerrors} shows a 2-d representation of the probability of detecting a transient as a function
of both its flux and its duration.
In this plot we recover all the expectations we listed in Section~\ref{sec:known}.
It is evident that transients dimmer than the flux required to be detected in the best image (indicated by the lowest horizontal line)
have no chance to be detected.
Transients with a flux higher than the one required to be detected in any image (indicated by the
highest horizontal line), and with a duration between the maximum separation between two consecutive observations and the
length of the observing campaign (indicated by the two vertical lines on the right), are always detected as transients.
Sources that are longer than the survey duration and are bright enough to be always detected can be treated as constant
sources if they start before the first observation in the survey. The probability that this happens is proportional to their duration
and this explains the gradient to the right of the right-most vertical line.
Moreover, very short transients (much shorter than one observation) which have a very bright flux, so bright that they are still
detectable in one observation, have a probability of being detected equal to the ratio of the amount of the total amount of
observing time during the survey and the time between the beginning and the end of the survey itself.
In fact, in that region (flux above 30\,Jy and duration shorter 5\,min), the probability of detection is constant at a value of
$\sim0.03$, which matches the ratio between the total observing time and length of the survey.
Finally, for transients shorter than the duration of one snapshot the minimum flux they must have to be detectable is a function
of their duration. This is due to the fact that their flux is averaged over the duration of the observation, and to be detectable, the
product of the flux and duration of the transient (i.e., its fluence) has to be larger than the product of the sensitivity and the integration
time of the observation.
This is indicated by the diagonal edge to the coloured region left of the leftmost vertical line.
We will investigate this further in Section~\ref{sec:fluence}.

Apart from these asymptotic approximations, there are other regions in the flux-duration plane for which the behaviour of the probability
strongly depends on the observing setup and in order to calculate the probability of detection in these intermediate regimes, the full
simulations are required.
This applies, for example, to
transients longer than the observing campaign and with flux between the sensitivity of the best
and the worst image of the dataset: they are likely detected as transients and not treated as constant sources.
This is due to the fact that these sources are bright enough to be detected in some images, beyond the extra threshold, but also
dim enough to be missed in some others.
Another intermediate regime where simulations are required is for sources with duration between T$_{obs}$ and T$_{max\_gap}$.
In these regions the probability depends on the timing setup of the observing campaign and on the sensitivity of the individual
observations.
To further illustrate the behaviour of the probability and transient rate as a function of flux and duration, we show in the
bottom two panels of Figure~\ref{fig:noerrors} cross sections of the top panel at one given flux (53\,Jy) and one given
duration (52\,days); we discuss these examples in the next two subsections.

\subsection{Transient rate as a function of flux}
In the bottom left panel of Figure~\ref{fig:noerrors} we show the probability and the transient rate as a function of the flux of the
transients for sources with a duration of 52\,days.
The first feature in this plot is that at a certain value of the flux the first transients start to be detected and their number keeps
increasing: this value of the flux corresponds to the minimum flux that a source must have to be detected in the best image
of the dataset (indicated by the left-most line).
The second feature we see is that the probability of detecting transients increases until it saturates and
remains constant. The increase is due to the fact that increasing the flux allows sources to be detected in more and more
observations until they reach a point where they can be detected in every observation. The flux at which this happens is marked
with the right-most line; it is clear that this is not the flux at which the probability flattens.
This is due to the fact that when a new source is detected, the code checks if it could have been detected
in the previous best image applying a higher threshold than the one usually used. Only if this is the case the source
is marked as transient.
This is implemented in order to avoid spurious detections of constant sources whose flux is very close to the detection
threshold.
The flux at which the probability flattens is the one at which sources are brighter than the extra threshold applied to
the best image in the dataset and it is marked by the middle vertical line.

\subsection{Transient rate as a function of duration}
\begin{figure*}
\begin{center}
\includegraphics[scale=0.75,viewport=0 0 600 396,clip]{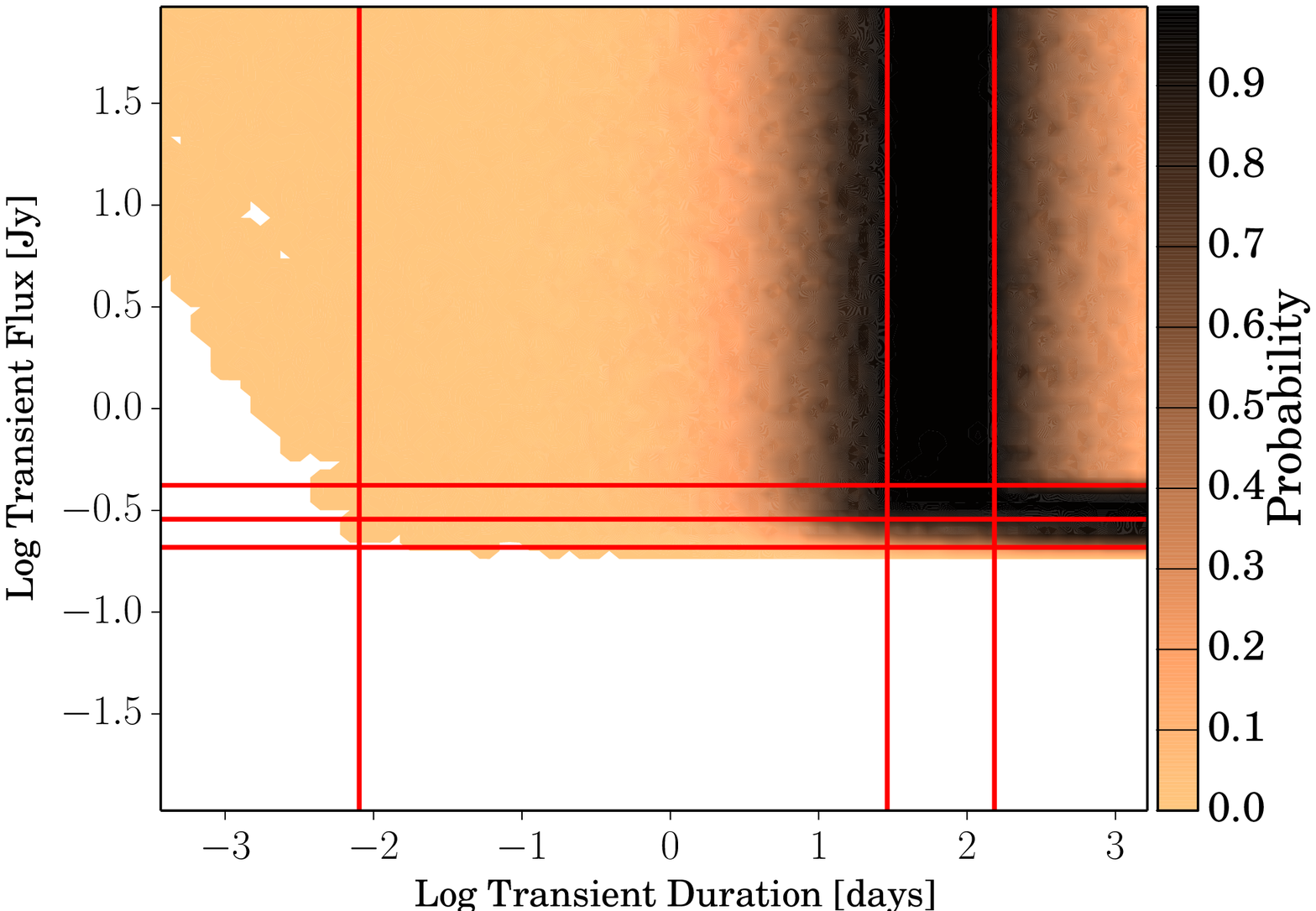}
\includegraphics[scale=0.4,viewport=0 0 585 410,clip]{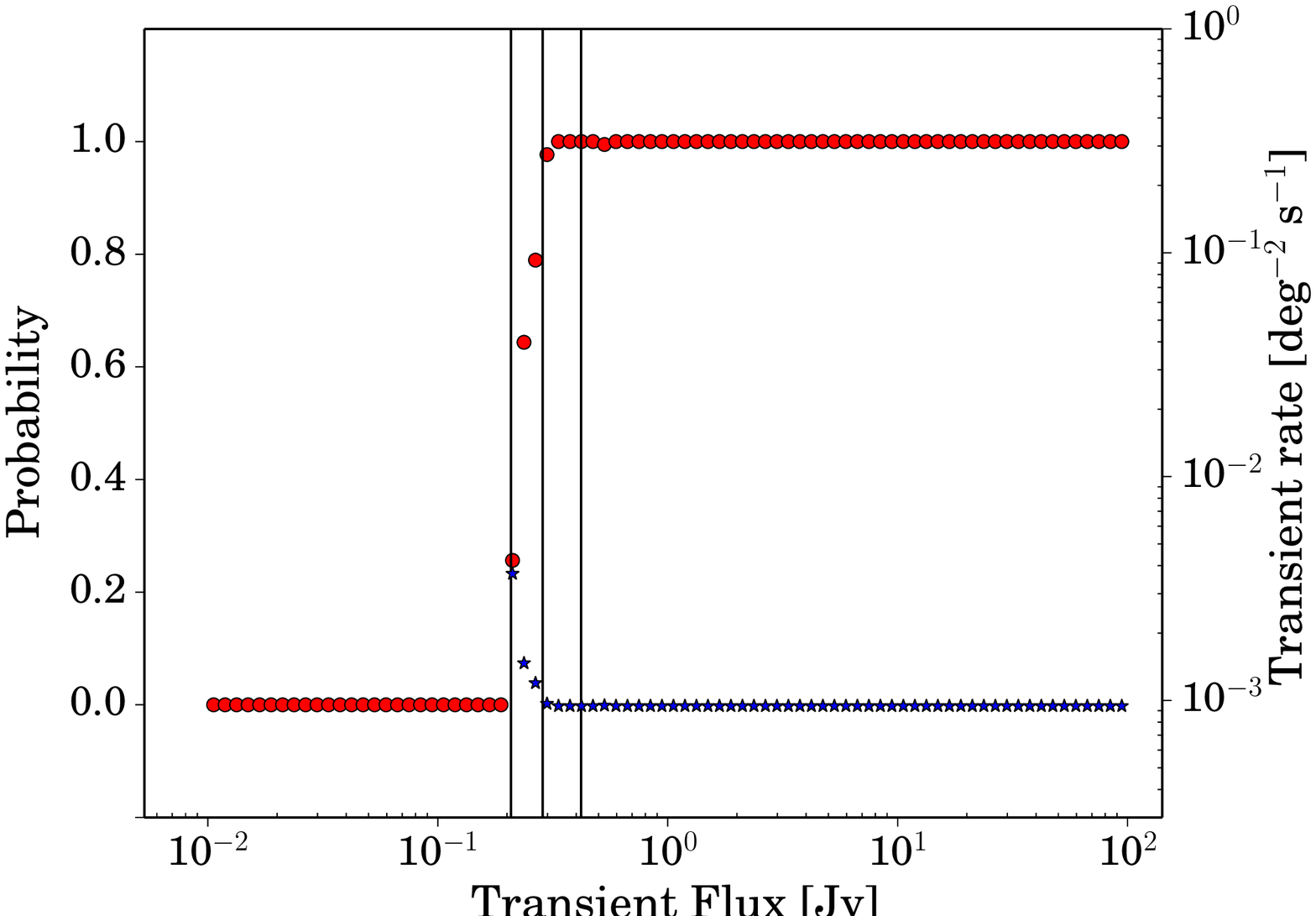}
\includegraphics[scale=0.4,viewport=0 0 585 410,clip]{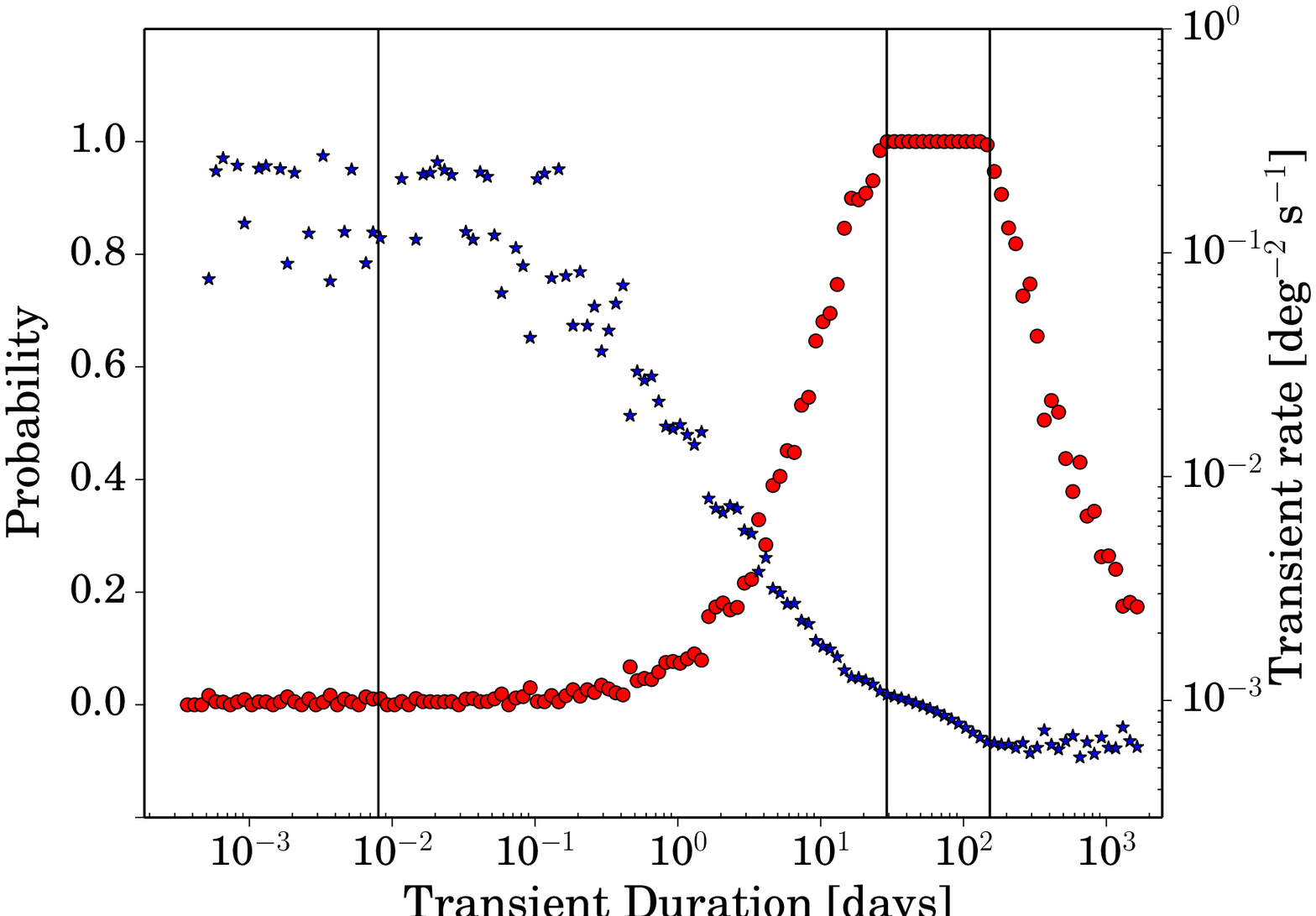}
\caption{Simulation of 2 million top-hat transients with no flux errors.
The top panel represents the probability for transient detection as a function of flux and duration.
The probability of detecting a transient source and the transient rate as a function of flux (for a duration of 52\,days) are shown
in the bottom left panel, and the probability and rate as a function of duration (for a flux of 53\,Jy) are shown in the bottom right.
In the latter two panels, the red dots represent the probability of detection whereas the blue stars represent the upper limits on
the transient rate.
In the bottom panels the values of the transient rate are not reported if the probability of detection is equal to zero;
this means that no constraints on the transient rate can be inferred.
The large scatter in the transient rate observed in the left part of the bottom right panel is due to fluctuations in the extremely low
values of the probability reflected in large scatter in the values of the transient rate which is inversely proportional to that. These
fluctuations are due to low number statistics in the detection of very short transients.
}
\label{fig:noerrors}
\end{center}
\end{figure*}

In the bottom right panel of Figure~\ref{fig:noerrors} we show the probability and the transient rate as a function of the duration of the
transients for sources with a flux of 53\,Jy.
The trend of the probability (and also of the transient rate) can be divided into four parts.
Short transients are detected very sporadically because the probability that they fall into gaps between observations is very
high or their flux within an observation is averaged to lower values that can make
them undetectable. This causes the flat start of the probability curve at values very close to zero. The probability starts rising
monotonically at durations of about 1\,hour, corresponding to the duration of the observations performed within the same day.
This time-scale is indicated by the left-most line in both the top and the bottom right panel of Figure~\ref{fig:noerrors}.
The probability continues to increase until it saturates at a time-scale of about 30\,days. This time-scale corresponds to the longest
gap between two consecutive observations, which means that transients with a longer duration will always be detectable at least
one observation and that their detectability depends only on their flux. This time-scale is marked with the second line in the bottom
right panel.
The rise in probability is followed by a plateau until time-scale of about 150\,days, corresponding to the total length of the
observing campaign (indicated by the right-most line). Transients longer than this time-scale will start being detected in every
observation and therefore being flagged as constant sources.
The constant decline on the right hand of the plateau is due to the fact that transients can start one duration before
the beginning of the observing campaign.
The large scatter in the transient rate observed in the left part of the bottom right panel of Figure~\ref{fig:noerrors} is due to fluctuations in the
extremely low values of the probability reflected in large scatter in the values of the transient rate which is inversely proportional
to that. These fluctuations are due to low number statistics in the detection of very short transients.

\section{Realistic situation}
\label{sec:real}
\begin{figure*}
\begin{center}
\includegraphics[scale=0.75,viewport=0 0 600 396,clip]{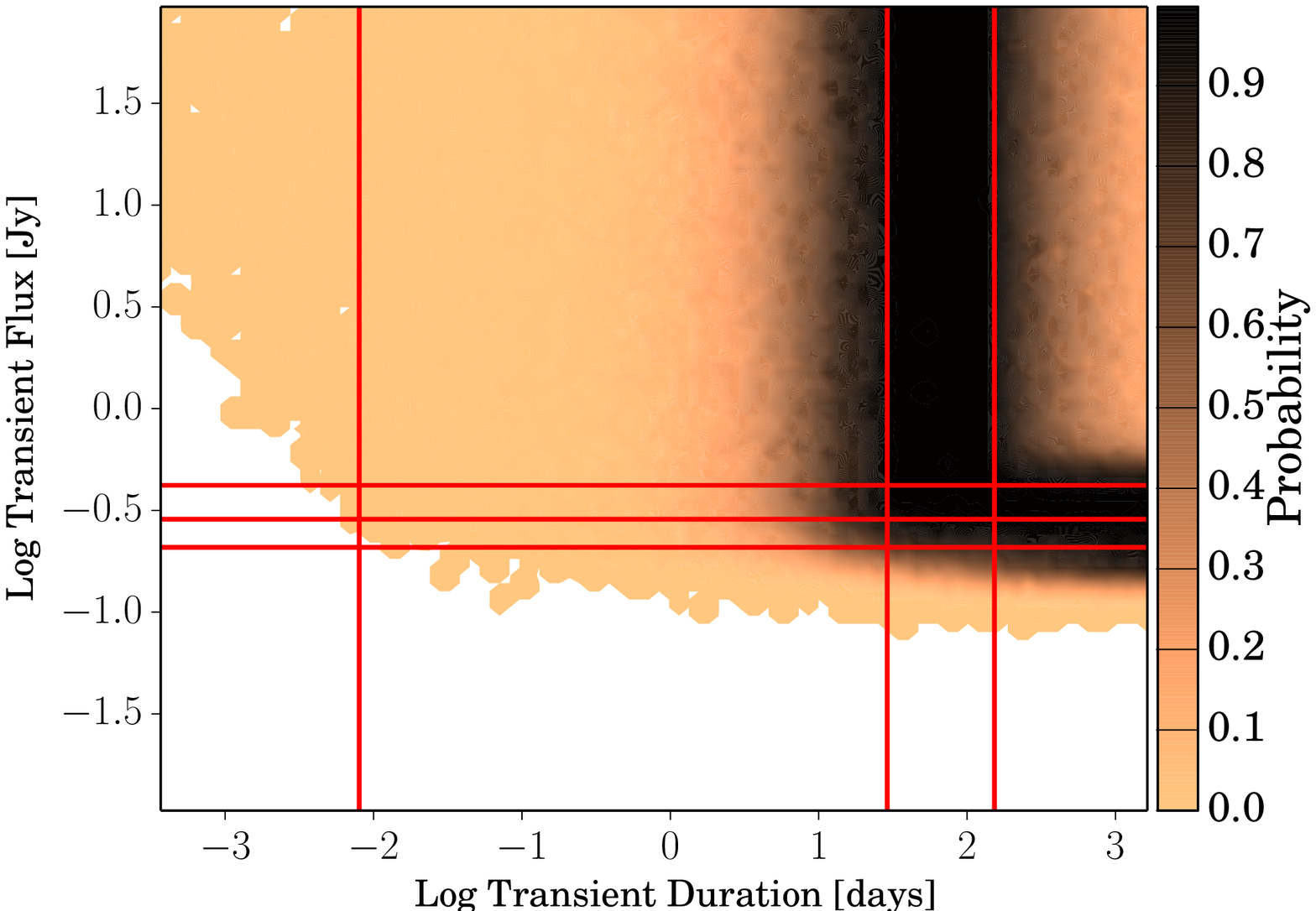}
\includegraphics[scale=0.4,viewport=0 0 585 410,clip]{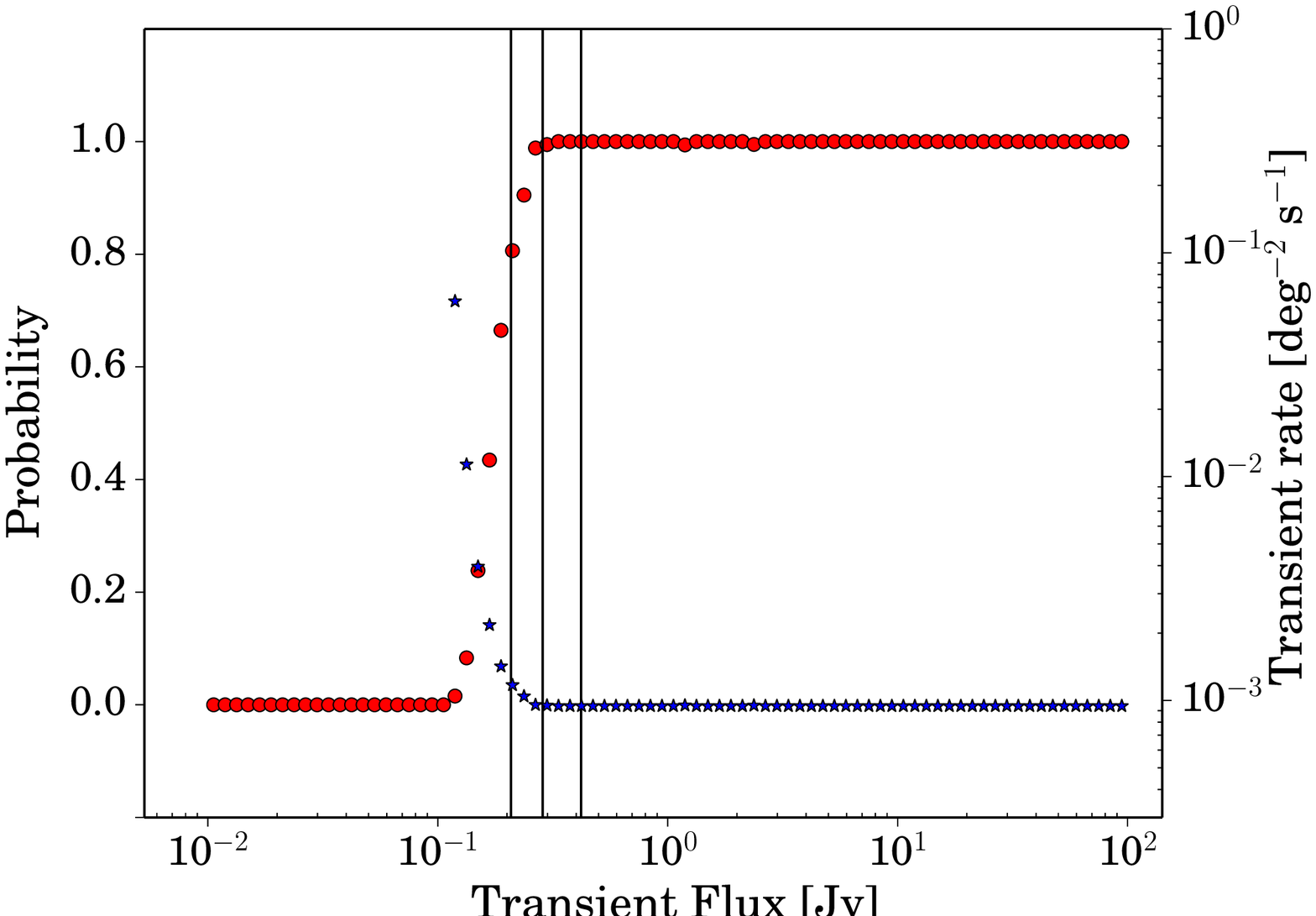}
\includegraphics[scale=0.4,viewport=0 0 585 410,clip]{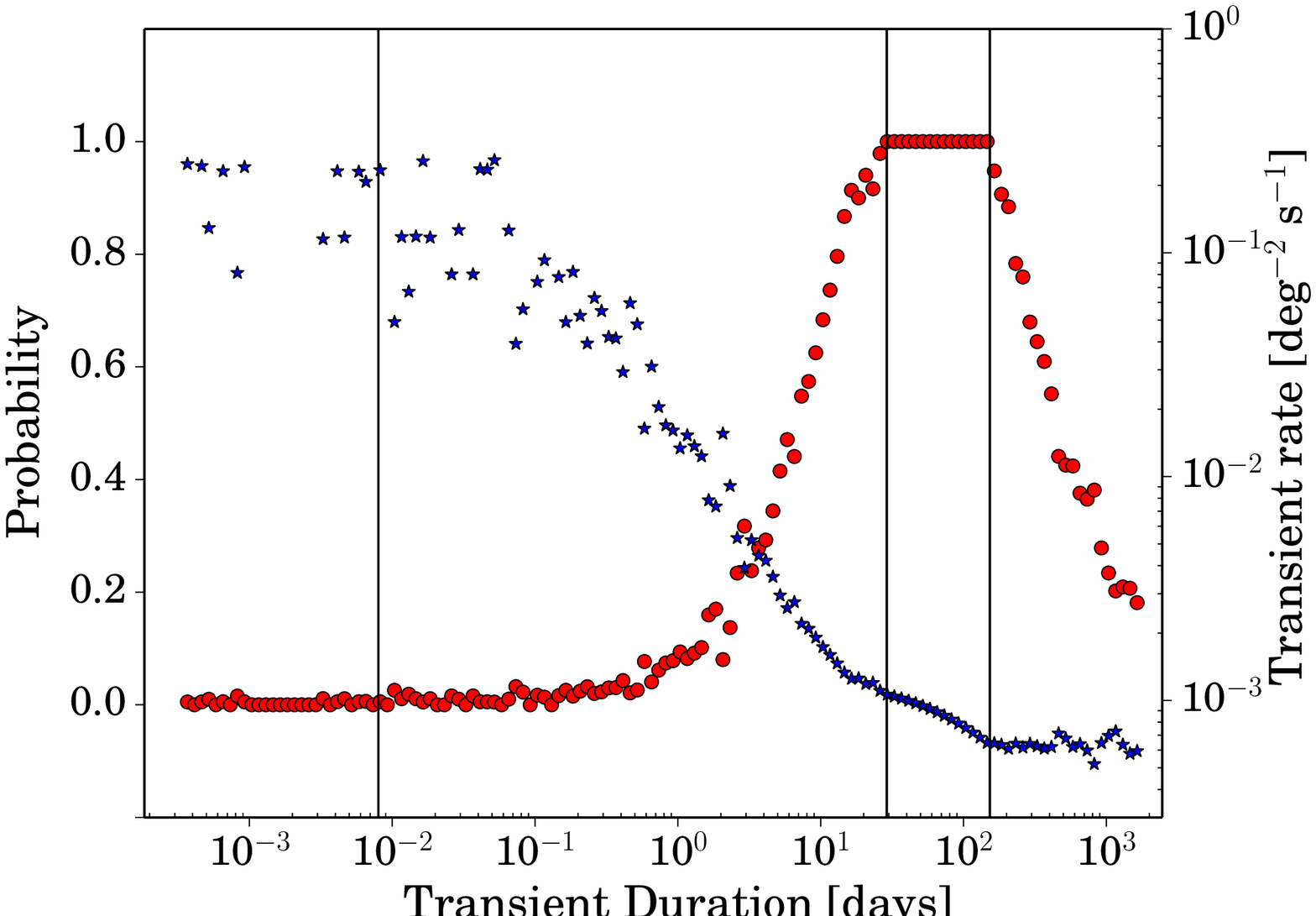}
\caption{The top panel represents the probability for transient detection as a function of flux and duration.
The probability of detecting a transient source and the transient rate as a function of flux (for a duration of 52\,days) are shown
in the bottom left panel, and the probability and rate as a function of duration (for a flux of 53\,Jy) are shown in the bottom right.
In the latter two panels, the red dots represent the probability of detection whereas the blue stars represent the upper limits on
the transient rate. In the bottom panels the values of the transient rate are not reported
if the probability of detection is equal to zero; this means that no constraints on the transient rate can be inferred.
The large scatter in the transient rate observed in left part of the bottom panels is due to fluctuations in the extremely low values of the
probability reflected in large scatter in the values of the transient rate which is inversely proportional to that. These fluctuations are
due to low number statistics in the detection of very short transients and in very faint ones.
These figures refer to top-hat function transients.}
\label{fig:tophat}
\end{center}
\end{figure*}

\begin{figure*}
\begin{center}
\includegraphics[scale=0.75,viewport=0 0 600 396,clip]{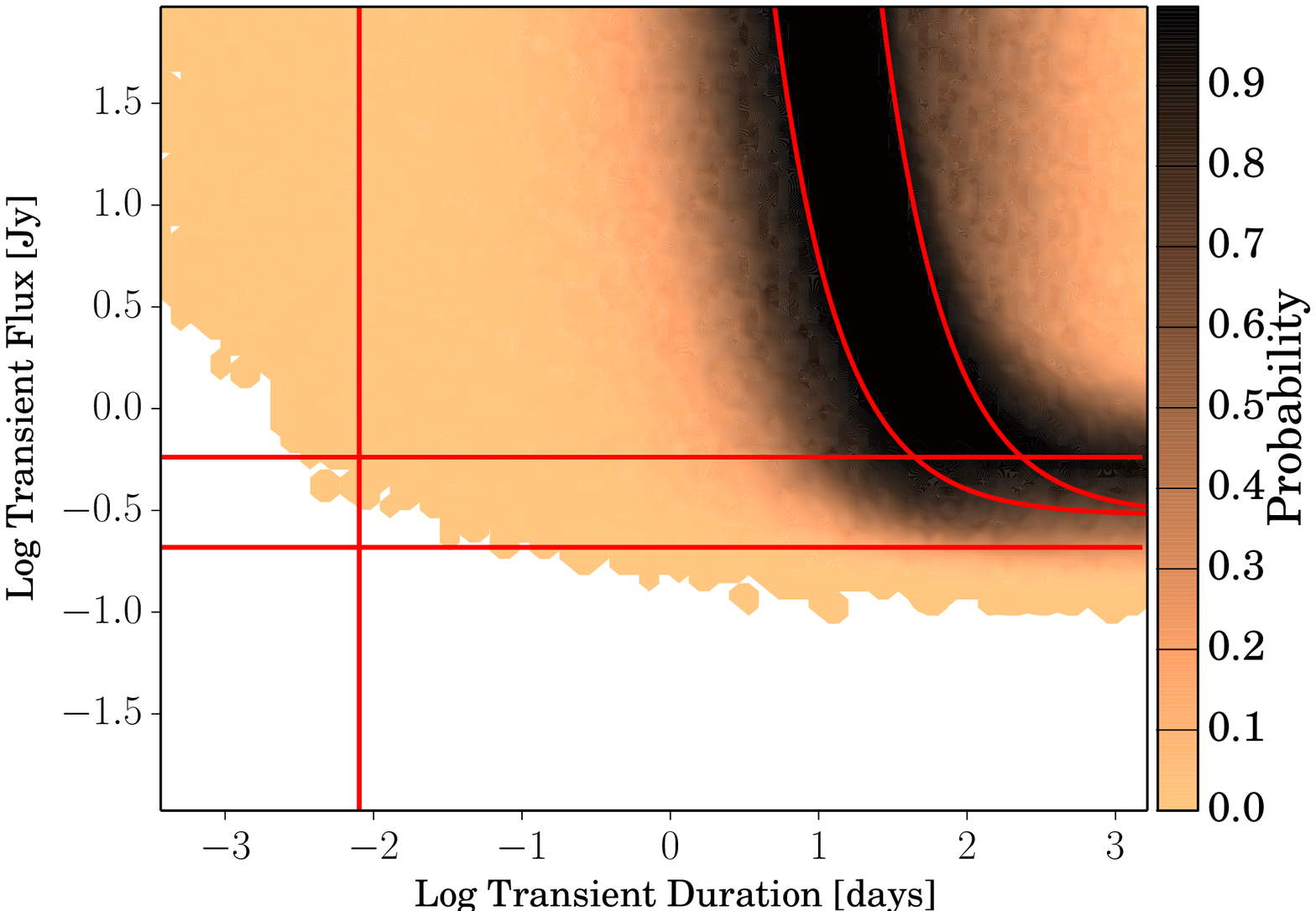}
\includegraphics[scale=0.4,viewport=0 0 585 410,clip]{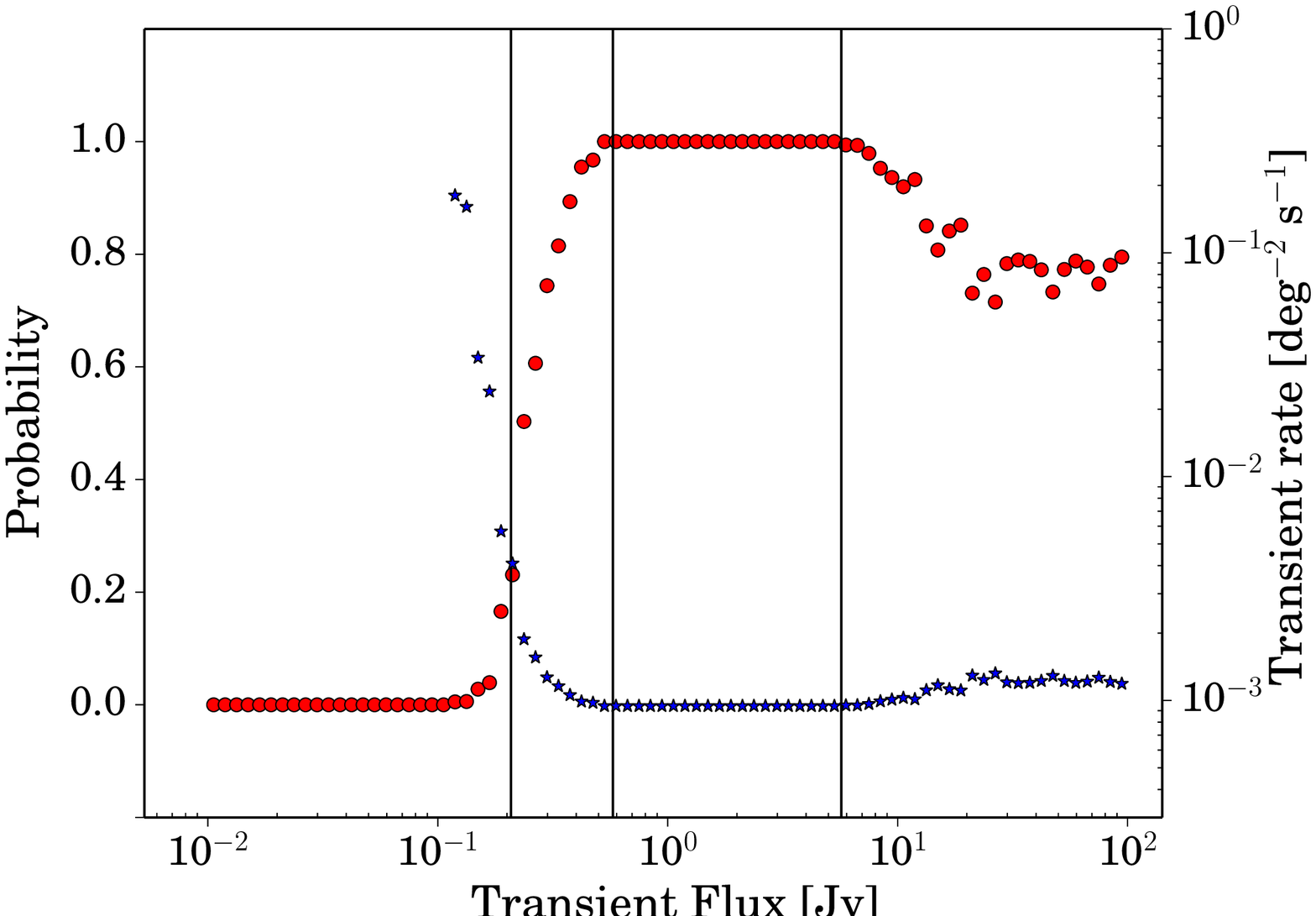}
\includegraphics[scale=0.4,viewport=0 0 585 410,clip]{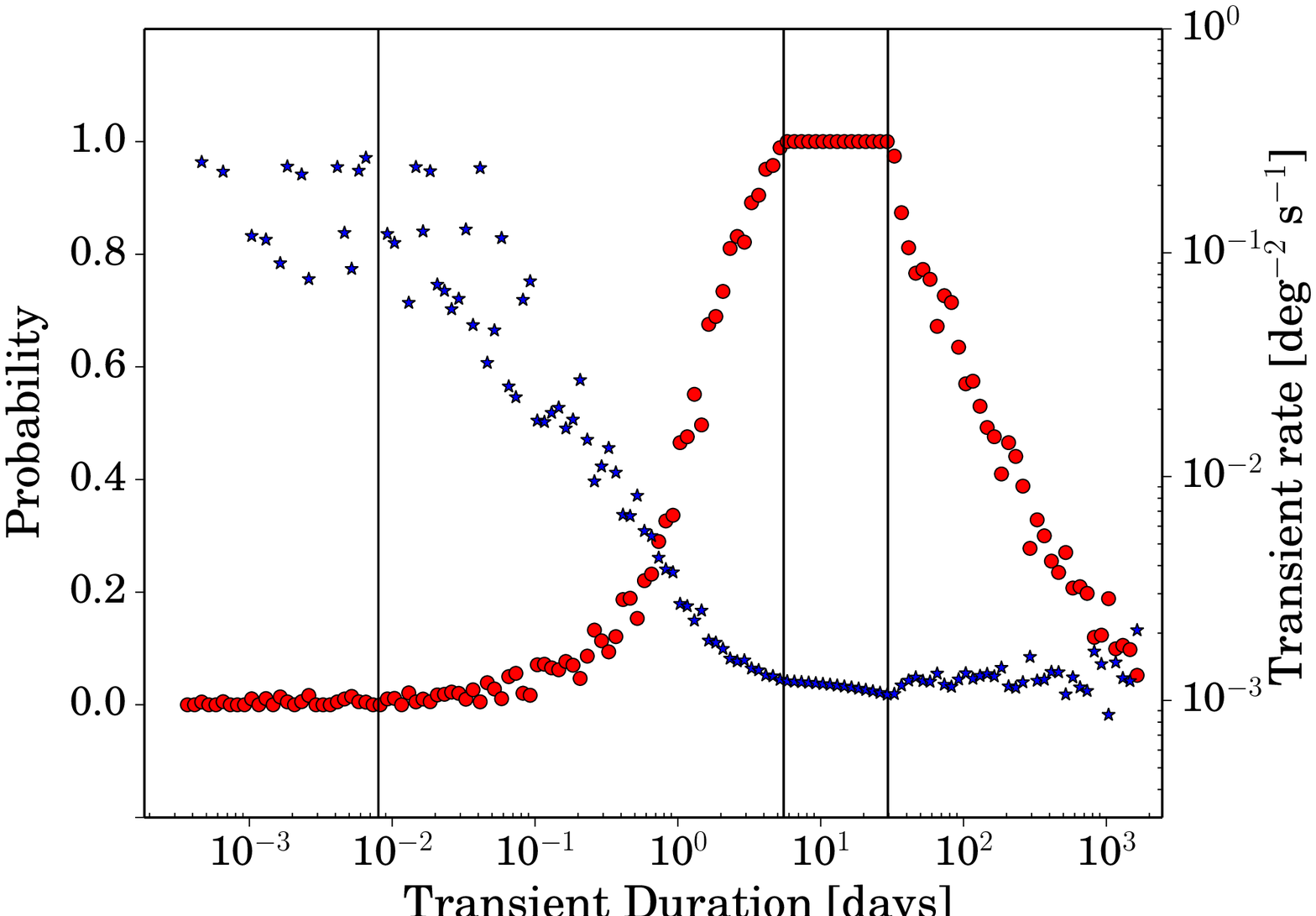}
\caption{
The top panel represents the probability for transient detection as a function of flux and duration.
The probability of detecting a transient source and the transient rate as a function of flux (for a duration of 52\,days) are shown
in the bottom left panel, and the probability and rate as a function of duration (for a flux of 53\,Jy) are shown in the bottom right.
In the latter two panels, the red dots represent the probability of detection whereas the blue stars represent the upper limits on
the transient rate. In the bottom panels the values of the transient rate are not reported
if the probability of detection is equal to zero; this means that no constraints on the transient rate can be inferred.
The large scatter in the transient rate observed in the left part of the bottom right panel is due to fluctuations in the extremely low values of the
probability reflected in large scatter in the values of the transient rate which is inversely proportional to that. These fluctuations are
due to low number statistics in the detection of very short transients.
The vertical lines in the bottom left panel represent, from left to right, the sensitivity of the best observation, the sensitivity of the worst image
including an extra threshold, and the solution of Equation~\ref{eq:fred_maior} for a duration of 52\,days.
The vertical lines in the bottom right panel represent, from left to right, the duration of one observation, and the solutions of
Equations~\ref{eq:fred_minor} and \ref{eq:fred_maior} for a flux of 53\,Jy.
These figures refer to FRED function transients.}
\label{fig:fred}
\end{center}
\end{figure*}

After validating our simulations code, we then applied it to a more realistic situation.
In this case we take into account errors in the flux measurements caused by, e.g., thermal noise and calibration errors.
In our simulations we included calibration errors by adopting a fractional error on the flux of each transient.
The value of this error is drawn from a gaussian distribution with mean value equal to 20\% and standard deviation equal to 5\%.
We also included a scatter in the flux due to the background noise in the observations. This contribution is different for
different observations and is equal to the absolute value of the noise level.
These two effects were added in quadrature to obtain the final value of the error in each observation.
The value of the flux of a source in an observation is then drawn from a gaussian distribution with central value equal to its
nominal flux and standard deviation equal to the total error we described.
This implies that each source has a different flux at each observation, as would also be the case in real observations.
In this simulations we used the same setup and method as in idea case presented in Section~\ref{sec:noerrors}.
Here we show the results of our simulations with both top-hat (Figure~\ref{fig:tophat}) and FRED (Figure~\ref{fig:fred})
transients, using the same flux and duration distributions.

The top panels in Figure~\ref{fig:tophat} and \ref{fig:fred} show the probability of detecting a transient as a function of both
its flux and its duration for top-hats and FREDs respectively.
We can see that top-hats show the same general trend as described in Section~\ref{sec:noerrors}.
The main difference with the top panel of Figure~\ref{fig:noerrors} is that long sources fainter than the sensitivity of the best observation
can now be detected whereas they were not in the ideal case.
This is due to the fact that these sources should be too faint to be detectable (in fact the lowest
horizontal line in the top panel of Figure~\ref{fig:noerrors} was a very sharp boundary for detections), but the scatter in flux makes it
possible for them to have a flux that is high enough to be detected in at least one observation. This effect is more prominent for longer
durations because they are on during more and more observations and therefore they have more trials and thus a higher probability
than short ones.
Also FREDs show a region of the top panel of Figure~\ref{fig:fred} in which the probability is equal to 1, but in this case the boundaries
of this region are flux dependent.
This is due to the fact that the fluxes of these sources do not drop to zero instantaneously after the time we called ``duration''.
This means that for very bright sources, the amount of flux being emitted after one duration is enough for them to be still detectable after that time.
The lower boundary of the region where sources are detected as transients for sure is set by the flux a source of a certain duration,
starting at the beginning of the longest gap between two consecutive observation must have to be still detectable in the following observation.
The function describing it is:

\begin{equation}
S(T)\,=\,S_{obs} \frac{T_{obs}}{T} \left(e^{-T_{max\_dur}\,/\,T}\,-\,e^{-(T_{max\_dur}\,+\,T_{obs})\,/\,T}\right)^{-1} \: .
\label{eq:fred_minor}
\end{equation}

\noindent This boundary is indicated by the leftmost curve in the top panel of Figure~\ref{fig:fred}.
The upper boundary of that region is set by the peak flux a source of a duration T, starting one duration before the beginning of the first
observation (the earliest possible start time) must have to be detectable by the last observation.
The function describing it is:

\begin{equation}
S(T)\,=\,S_{obs} \frac{T_{obs}}{T} \left(e^{-(T_{survey}\,+\,T\,-\,T_{obs}) / T}\,-\,e^{-(T_{survey}\,+\,T)\,/\,T}\right)^{-1} \: .
\label{eq:fred_maior}
\end{equation}

\noindent This boundary is indicated by the rightmost curve in the top panel of Figure~\ref{fig:fred}.
We can see that the region where the probability of detection is 1 is also delimited by the highest horizontal line, 
representing the flux of transients that are detectable in any observation including the extra threshold.
This is due to the fact that sources fainter than that are not always detectable in all the observations.

Comparing the bottom right panels in Figure~\ref{fig:tophat} and \ref{fig:fred}, we see that the overall trend is the same.
The probability is close to zero at the beginning, and then it starts rising around the duration of one observation; it flattens
reaching the region where the probability is equal to one and rapidly drops after its end.
The difference between the top-hats and FREDs is the timescale at which the probability saturates at 1 and after which it starts decreasing.
These timescale are shorter for FREDs and this again is due to the fact they have a tail of flux extending above their nominal duration.

When comparing the trend of the probability as a function of the transient flux (bottom left panels of Figure~\ref{fig:tophat}
and \ref{fig:fred}), we can see that top-hat transients start being detected with fluxes smaller than FREDs, and in both
cases at a flux lower than the minimum required to be detected in the best image. The fact that sources start being
detected even if their flux should not be high enough, is explained by the flux errors; in some images
the flux of such sources is high enough for a detection, also above the extra threshold.
The fact that top-hat sources are detected at lower fluxes than FREDs is due to the same reason why very long and faint sources are
detected if they are top-hat but not if they are FRED.
Another notable difference between the bottom left panel of Figures~\ref{fig:tophat} and \ref{fig:fred} is that at very high flux the
probability of detection drops from the value of 1. This is due to the fact that these sources start to be treated as constant because
the flux contained in their tail is high enough for them to be detectable way above their nominal duration.

\section{Discussion}
\label{sec:discussion}

\subsection{Fluence is what matters}
\label{sec:fluence}
For a transient to be detected in an image the most relevant quantity is the amount of flux falling within the 
integration time of a given observation, i.e., the fluence in an observation. This quantity can be high both because the
source is very bright for a short time or because it has a long duration.

\begin{figure*}
\begin{center}
\includegraphics[scale=0.43,viewport=0 0 560 410,clip]{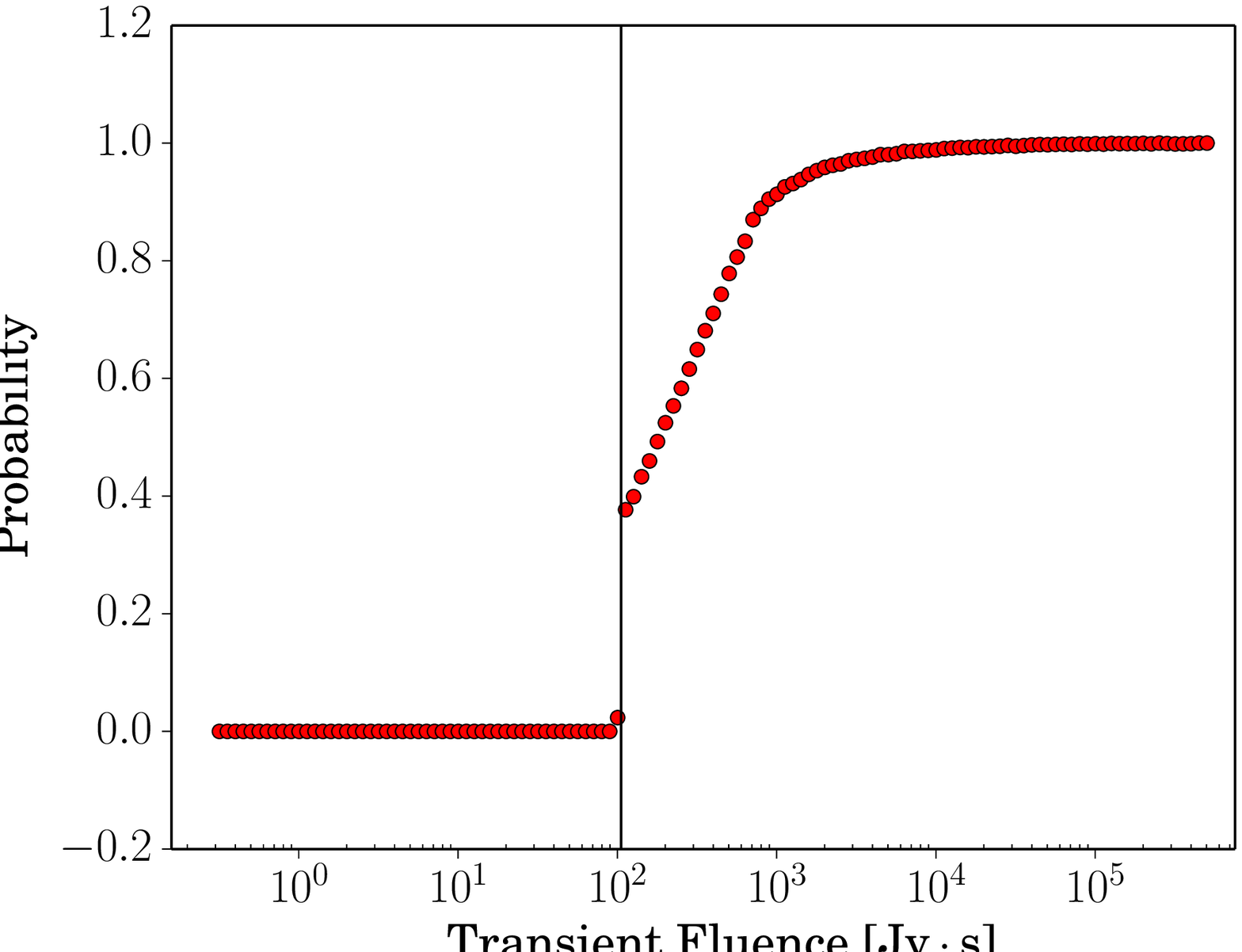}
\includegraphics[scale=0.43,viewport=0 0 527 410,clip]{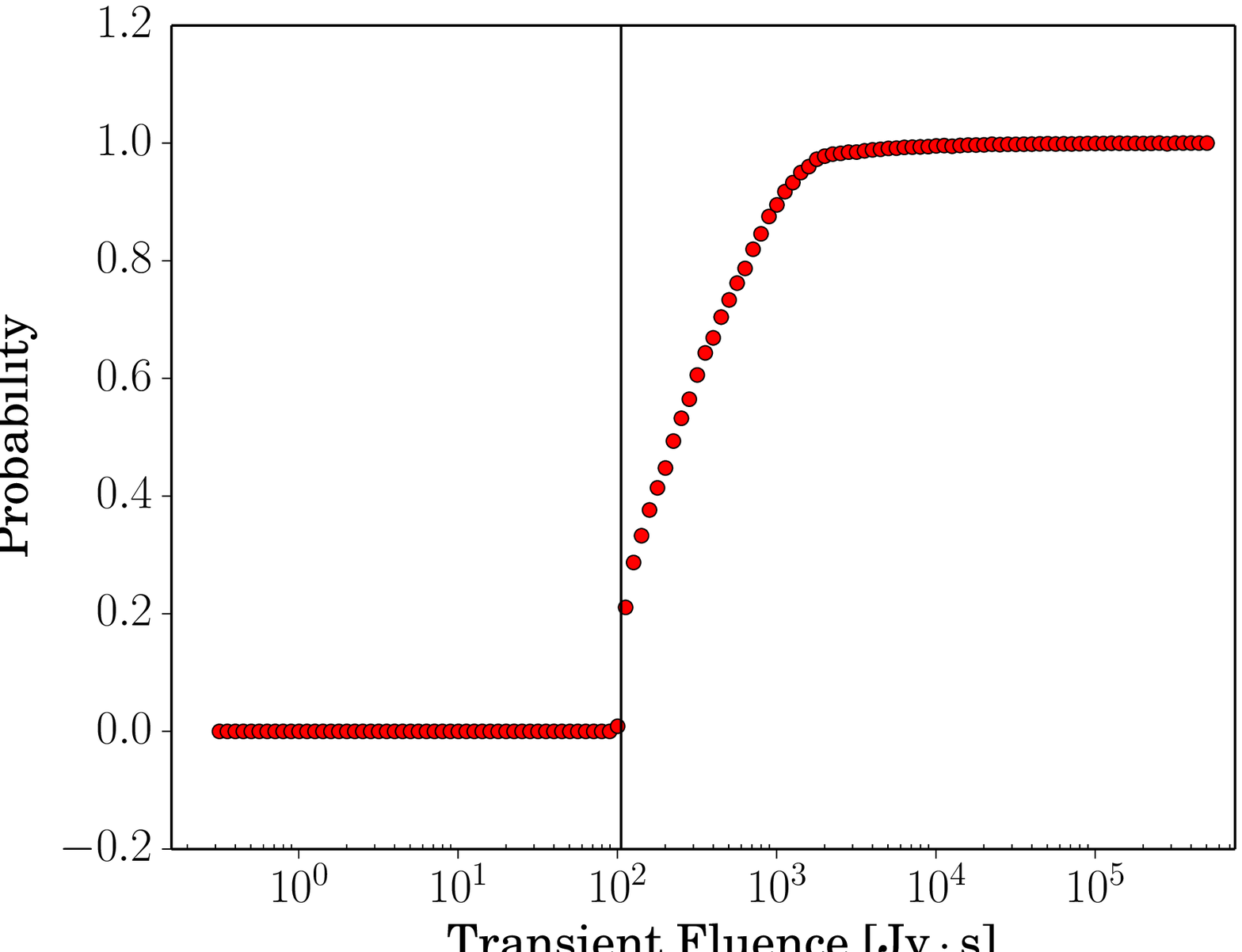}
\caption{Simulation of 2 millions transient sources with a top-hat lighcurve on the left and with a FRED lightcurve
on the right. The plots show the probability of a transient to be detected as a function of its fluence.}
\label{fig:fluence}
\end{center}
\end{figure*}

To show this we have taken one single observation and simulated both top-hat and FRED lightcurve transients, starting
between one duration before the beginning of that observation and the end of it. For clarity, we have imposed no errors
on their fluxes.
The results of these simulations are shown in Figure~\ref{fig:fluence} where we plot the detection probability as a function
of the fluence of the transients (which is the same as or larger than the fluence that falls within the observation).
The panel on the left shows the results we get for top-hat transients while the one on the right is for FREDs.
We expect that only transients with a fluence higher than the sensitivity of the observation multiplied by its duration (indicated
by the line in both panels of Figure~\ref{fig:fluence}) can be detected.
We can see that sources which have a fluence larger than the sensitivity of the observation multiplied by its duration can still
be missed as they might start too early or too late, and their fluence within the observation might be too small.

Figure~\ref{fig:fluence} shows that whenever a transient has a fluence larger than the sensitivity of the observation multiplied
by its duration, it is more probable to be detected when it has a top-hat lightcurve rather than a FRED. This is due to the fact
that if the peak of the FRED falls out of the observation, large part of the fluence is missed. On the other hand, if the fluence is
much higher than the sensitivity of the observation multiplied by its duration, the detection probability of FREDs  saturates at
one before it does for top hat. 
This happens because transients in this regime have both high flux and high duration: 
the flux of top hat transients will drop to zero instantaneously and can therefore be missed if their duration
inside the observation is not long enough; whereas the flux of FREDs will extend longer, and as the initial flux is high and duration long,
the flux in the tail will still be enough for the transient to be detected.

More generally, for transients shorter than one observation, the flux limit for detection is not set by
the sensitivity of the observation, but it depends on the duration of the transient as well.
Specifically, the fluence of a transient in an observation (flux $\times$ duration in the observation) has to be larger than the product of the
sensitivity of the observation and its duration. For a top-hat shape transient this expression can be easily written as:
S\,$\ge$\,S$_{obs}\,\times\,\frac{T_{obs}}{T}$; where S and T are the flux and the duration of the transient while S$_{obs}$
and T$_{obs}$ are the sensitivity and the integration time of the observation. This effect can also be seen in the top panels of
Figures~\ref{fig:noerrors}, \ref{fig:tophat} and \ref{fig:fred}.

\subsection{Comparison with other methods}
\label{sec:comparison}
Many works have explored the transient sky and developed methods to calculate or constrain the transient rate
(or surface density).
In most of the cases \citep[e.g.,][]{VLAarchival_Bell2011, MOSTtrans_Bannister2011}
the authors limited themselves to constraining the transient rate (or surface density) as a function of
the flux of the transients. They assumed an isotropic population of transients across the field of view they surveyed, and
calculated the total area they monitored to derive their limits.
\citet{Croft2013, MWAtrans_Bell2014, PanSTARRS_Carbone2014, transsel_Rowlinson2016, transients_ofek2011}
modelled the variable noise across their field of view and calculated the surveyed area as a function of the sensitivity to constrain
the transient rate for multiple values of the flux of the transients.
In a few cases \citep[e.g.,][]{Adam_Stewart2015, transsel_Rowlinson2016, vlite_Polisensky2016}
the time dimension has been explored as well, in order to constrain the transient rate (or surface density) as a function of the
duration of the transients.
In none of these works the three dimensions (transient rate, flux, and duration) have been accounted for systematically and rigorously.
This and the fact that we can populate the whole flux-duration plane with value or upper limits to the transient rate are the main
advantages of our simulations method.

In a previous work \citep{PanSTARRS_Carbone2014} we analysed the relationship between the transient
rate ($\hat{\rho}$) and the duration of the transients ($T$).
In that paper we were sensitive to time-scales ranging from 15 minutes to about 5 months.
We assumed a Poisson distribution to calculate the 95\% confidence level upper limit to the transient rate, as in
Equation~\ref{eq:poisson}.
We rewrote the Poisson variable $\lambda$ as the product of the transient rate, the total
number of square degrees, and the total observing time: $\lambda$\,=\,$\hat{\rho}$\,$\Omega_{tot}$\,T$_{tot}$\,.
This gives us the number of transients for a given time-scale falling within an observation (n$_{\textrm{obs}}$).
The total surveyed area can be written as the product of the field of view of every observation
(if this is constant) and the number of different fields being monitored (in this case it is 1).
The total observing time is a function of the timescale of the transients because only statistically independent pairs of
observations add information: T$_{tot}$(T)\,=\,N$_{pairs}$(T)\,\,T$_{obs}$, where T$_{obs}$ is the integration time of each
observation.
To calculate the number of independent pairs of observations separated by a time $T$, we binned together images with a time
difference which was lower than the time-scale of interest.
For example, all the observations from the same day collapse into one measurementt for time-scales longer than a day,
two observations within a week are merged for time-scales longer longer than a week, and so on.
If no transients were detected ($n$\,=\,0) we can set upper limits the transient surface density at the 95\% confidence level as:

\begin{equation}
\hat{\rho}_{obs}(T)\,<\,-\frac{\textrm{ln}(0.05)}{\Omega\,\times\,N_{\textrm{pairs}}(T)\,\times\,T_{obs}} \ .
\label{eq:rate_timescale_classical}
\end{equation}

\begin{figure*}
\begin{center}
\includegraphics[scale=0.8,viewport=0 0 567 396,clip]{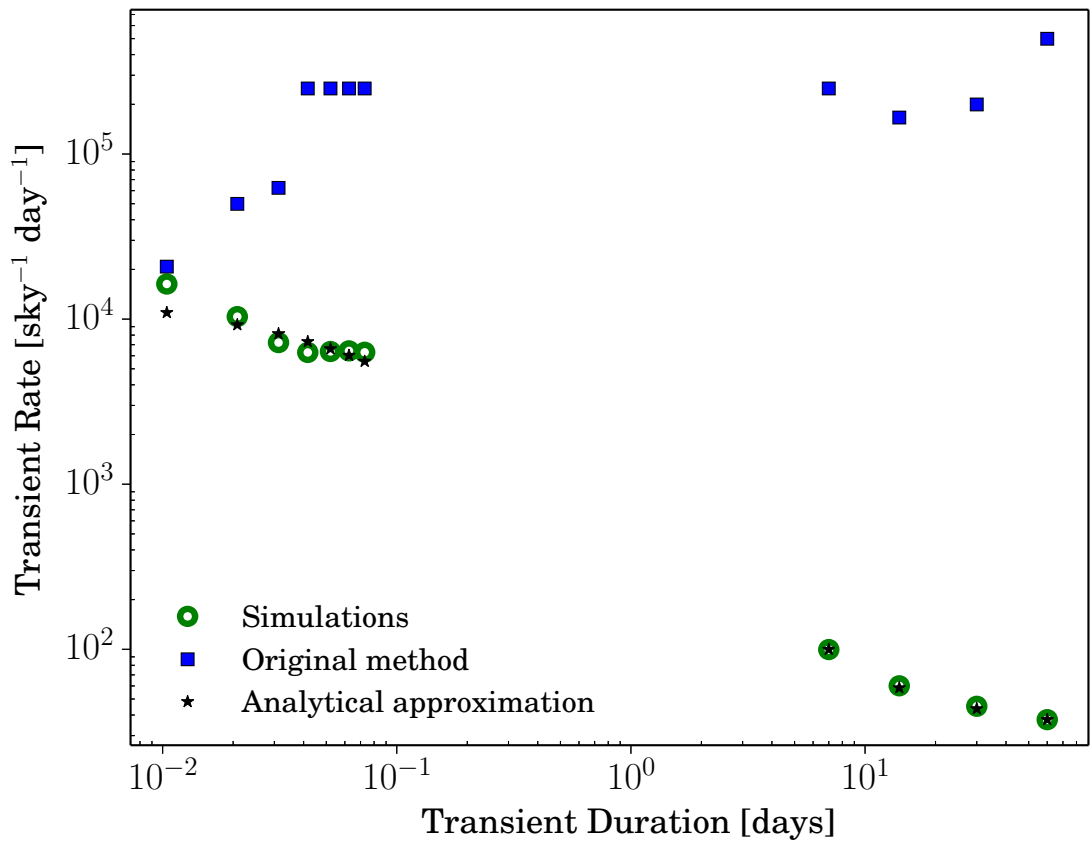}
\caption{Comparison between the methods from \citet{PanSTARRS_Carbone2014}
and this work in calculating the transient rate as a function of the duration of the transients. With blue squares
we plot the method from \citet{PanSTARRS_Carbone2014} that did not take into account gaps between observations;
with black stars we show the corrected method they suggested; and with green circles we show the
results from the simulations in this paper.
}
\label{fig:comparison}
\end{center}
\end{figure*}

\noindent The upper limits we get using this method are displayed in Figure~\ref{fig:comparison} as blue squares.

We can also calculate the transient rate dividing the total number of transients between the beginning and the end of
our survey by the total amount of square degrees we survey and the total amount of time we were sensitive to a transient
with duration T.
The total amount of time we were sensitive to a transient as a function of its duration 
is equal to the total survey time plus one duration, meaning that the transient could have started
one duration before the first snapshot and still be detectable in it
(T$_{tot}$(T)\,=\,T$_{survey}$\,+\,T).
To get an estimate of the total number of transients between the beginning and the end of
the survey, we have to correct for the probability of a source of duration T to fall completely
in gaps between two observations (p$_{\textrm{gaps}}$). This probability is equal to the ratio between the amount of time when
a transient can start without being detectable in any observation and the total duration of the survey.
A transient of duration T is detectable in an observation starting at $t_{\textrm{start}}$ and finishing
at $t_{\textrm{end}}$ if it starts between $t_{\textrm{start}}$\,--\,T and $t_{\textrm{end}}$. This implies that in a gap of length
$T_{\textrm{gap}}$ between two consecutive observations the amount of time when a transient of duration T can start and
not fall in any of the two observations is equal to max[($T_{\textrm{gap}}$\,--\,T), 0]. Summing this on all the gaps and dividing
by the total survey time we obtain the probability we were looking for:

\begin{equation}
p_{\textrm{gaps}}(T)\,=\,\frac{\sum_{\textrm{i}}\,\textrm{max}\left[(T_{\textrm{gap,\,i}} - T),\:0\right]}{T_{\textrm{survey}}} \ ,
\end{equation}

Now we can calculate the total number of transients between the beginning and the end of the survey (n$_{\textrm{tot}}$) as:

\begin{equation}
n_{\textrm{obs}}(T)\,=\,n_{\textrm{tot}}(T)\,\times\,(1\,-\,p_{\textrm{gaps}}(T)) \ ,
\end{equation}

The total number of transients can be expressed as
\newline$n_{\textrm{tot}}=\,\hat{\rho}_{\textrm{tot}}\,\Omega\,(T_{survey}\,+\,T)$
while the observed number of transients is given by
$n_{\textrm{obs}}=\,\hat{\rho}_{\textrm{obs}}\,\Omega\,N_{pairs}\,T_{obs}$. Therefore:

\begin{equation}
\hat{\rho}_{\textrm{tot}}(T)\,=\,\hat{\rho}_{\textrm{obs}}(T)\frac{N_{pairs}\,\times\,T_{obs}}{T_{survey}\,+\,T}\frac{1}{1\,-\,P_{\textrm{gaps}}(T)} \ ,
\end{equation}

\noindent where $\hat{\rho}_{\textrm{obs}}$ is given in Equation~{\ref{eq:rate_timescale_classical}} and can be used to derive:

\begin{equation}
\hat{\rho}_{\textrm{tot}}(T)\,<\,-\frac{\textrm{ln}(0.05)}{\Omega\,\times\,(T_{survey}\,+\,T)}\frac{1}{1\,-\,P_{\textrm{gaps}}(T)} \ .
\label{eq:timescales_newmethod}
\end{equation}

\noindent The upper limits we get using this method are displayed in Figure~\ref{fig:comparison} as black stars.

We compared the results from these two methods with the results of our simulations.
In order to make this comparison we binned the results from our simulations as in Section~\ref{sec:real} to the same time intervals as in 
\citet{PanSTARRS_Carbone2014}. The results from the simulations are represented as green circles in Figure~\ref{fig:comparison}.

It is evident how the results from the simulations are very well approximated by the second method described in
It is evident how the results from the simulations are well approximated by the second method described in
\citet{PanSTARRS_Carbone2014}, while the first method can put much looser constraints, especially for longer timescales.
This is due to the fact that both the simulations and the second method of \citet{PanSTARRS_Carbone2014}
take into account the fact that a transient can start before an observation and still be detectable
if at least part of it fall within the observation, while the first method of \citet{PanSTARRS_Carbone2014} only takes into account
transients starting within an observation.
The discrepancy is larger for longer timescales because the amount of time before an observation a transient can start and still be on
when the observation begins is equal to one transient duration.

The results from this work and the second method of \citet{PanSTARRS_Carbone2014} are different only at very short timescales.
This discrepancy is due to the fact
that the simulations take into account the fact that a transient that is on for at least part of the observation is not necessarily detected;
the probability that it is detected depends on a combination of its flux and its duration within the observation.
This problem affects mainly short duration transients because for longer duration ones, even if they start at the end of one observation,
the following one might still detect it.

\subsection{Power and benefits of the method}
The simulations presented in this paper only assume that the observations are of the same field, and that
the noise in the images is uniform; there are no further constraints on the setup of the observations and survey.
This means that this method can be used for surveys at any observing frequency, from radio to $\gamma$-rays.
In fact, Carbone \& Wijnands (in preparation) will use the method described in this paper to study the detectability
of Low Mass X-ray Binaries outbursts in X-ray surveys.
This method is not restricted to any specific source type, as long as the light curves of the transients can be modelled with the
given functions (other lightcurve shapes can be implemented easily).

Another important feature of this method is that it can be used for multiple aims.
If one wants to set up a survey looking for a specific type of object for which we roughly know the transient rate, this method
can foresee the average number of sources that such a survey will detect per flux and duration bin. This can help setting up and
optimising the parameters required for the survey, such as the separation between observations and the sensitivity.
Turning this argument around, once such a survey has been performed this method can help updating and correcting the
transient rate.
Another natural application for this method lies in extracting as much information as possible from a survey.
Currently, for most surveys a limited number of points in the flux-duration plane are calculated, whereas our method
allows to entirely fill the plane with values of the transient rate, or upper limits.

\subsection{Application to another survey}
\label{sec:survey_mwa}
\begin{figure}
\begin{center}
\includegraphics[scale=0.43,viewport=0 0 600 396,clip]{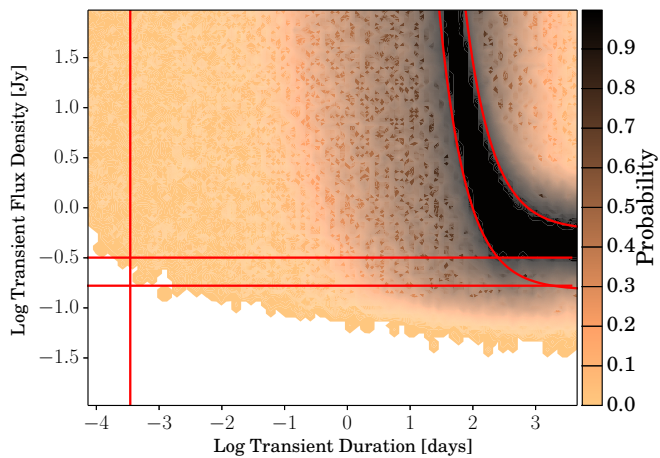}
\caption{
Simulation of 2 millions transient sources with a FRED lighcurve on the survey setup used for MWA EoR search
described in Section~\ref{sec:survey_mwa}. The same pattern as in the simulation from a different survey (see Figure~\ref{fig:fred})
is clear here, except that the features we can observe are shifted and stretched due to the different timescales and sensitivities
of this survey.
}
\label{fig:mwa}
\end{center}
\end{figure}

We have applied this simulation method to a different survey from a different instrument as well to demonstrate its 
flexibility.
The survey we simulate is the one described in \citet{transsel_Rowlinson2016}.
It was conducted using the Murchinson Widefield Array \citep[MWA;][]{MWA_tingay2013} between August 2013
and September 2014. Each observation was 28 seconds long and covered an area of 452\,deg$^2$.
The average noise of the images is 31\,mJy\,beam$^{-1}$. We have simulated 2 million FREDs-like transients,
including errors, and the results are shown in Figure~\ref{fig:mwa}.
The features we described in Section~\ref{sec:real} are also recovered in this case, when correcting for different values
of the survey parameters. 

This survey was designed to detect short radio transients, tens of seconds long, such as Fast Radio Bursts.
The probability that such short transients are detected is small because the survey design has long times of non observations
alternating periods with very dense observing coverage.
On the other hand, the number of expected events for transients of duration shorter than the
integration time of a single observation does not strongly depend on the presence of gaps
but only on the total amount of observing time of the survey.
Fast Radio Bursts are expected to have a very high whole sky rate \citep[see for example][]{Keane2015} and the aim was
to detect a handful of events among them.
This survey is very sensitive to transients of duration of tens of days thanks to the limited gaps between consecutive observations.

When comparing the limits on the Fast Radio Bursts all sky rate calculated by \citet{transsel_Rowlinson2016}
and the results that can be calculated from his work we obtain the same trend as for the dataset in
\citet{PanSTARRS_Carbone2014} discussed in Section~\ref{sec:comparison}.

\section{Future developments}
\subsection{Transient and variable sources}
In our simulations we do not distinguish between variable and transient sources.
We define as transients those variable sources for which the flux level in quiescence is well below the
sensitivity limit of the survey, and that undergo only one outburst during the simulated time interval.
A possible extension of the current simulation method to include variable sources with multiple outbursts 
is to include a new class of light curve mimicking the behaviour of traditional variable sources, e.g., a sinusoidal function.
Non periodic variables can also be simulated by associating different outbursts to the same source.
This technique will be applied in a follow-up work by Carbone \& Wijnands (in preparation) that will study the detectability
of Low Mass X-ray Binaries outbursts in X-ray surveys.

\subsection{Multiple pointings and variable noise floor}
We are aware that most transient surveys are not focused on a single field and have multiple pointings instead; moreover,
more realistic situations do not have uniform noise across the field of view.
We can easily extend the capabilities of our simulations to cope with more realistic scenarios by assigning spatial
coordinates to the simulated transients.
This way we can check if a transient lies in the field of view of an observation before checking whether it is detectable.
One caveat to this extension is that the source populations might be different towards different directions, e.g., galactic
vs extragalactic directions, and this must be taken into account.

We could take into account variable noise across the field of view of the observations in two ways.
We could divide the image into smaller ones where the noise is uniform and treat each of them as a separate pointing, or we could
model the noise variability across the field of view and measure the local noise at the location of the transient source.
The latter can be done assuming that the noise variability resembles the beam response as done in \citet{Croft2013} and used in 
\citet{transsel_Rowlinson2016}.

\subsection{Co-added map}
In many cases, transient surveys in the optical band, and sometimes at radio wavelengths, compare individual epochs with a
co-addition of all the epochs to find transients.
This technique does not apply to the simulations described in this work because we compare the sources detected in different epochs
with each other and not with an external catalogue, taking into account the time variable as well as the spatial coincidence.
Moreover, the method described in this paper has been developed to simulate the detectability of rare transient events
in long series of observations with limited sensitivity, whereas a combined reference map is very useful when many transients
and variable sources are present in the data, and understanding whether a transient is a new source or not is important
to retain only relevant events.

\section{Conclusion} 
\label{sec:conclusion}
We have presented a new method to analyse the result of a transients survey and translate the survey results as precisely
as possible into rates of transients as a function of flux and duration. 
While most previous efforts to estimate surface densities and rates are done with analytical approximations, our method involves
simulating transients of different fluxes and durations.
The main features of our simulations method are:
\begin{itemize}
\item Our method is independent of the frequency of the observations that are simulated.
\item Our method can also take into account
the type of sources that are simulated, as far as their light curve can be modelled
(e.g. a top-hat or FRED light curve).
\item For the first time, we can calculate the transient rate for every combination of flux and duration of transients.
\item We can easily convert the probability of detection in transient surface density and in transient rate, in the case
of detections as well as for upper limits in the case of non-detections.
\end{itemize}
To validate our simulations, we have given analytical approximations in certain asymptotic regimes of the flux-duration parameter
space, and performed simulations under ideal conditions in which the results can be easily interpreted.
The asymptotic regimes in which the analytical approximations are valid are limited, since the probability of detection is strongly
dependent on the survey setup; which is why simulations covering the full flux-duration plane are necessary.
In fact, we show that our method leads to significantly different transient rate estimates than previous efforts,
in particular in the transient rate as a function of transient duration, simply because there are effects that can not be
taken into account with analytical approximations.

Our method can be applied to surveys that are searching for specific types of source, and can help setting up the best survey
strategy to detect them. It can also be used to analyse surveys that have been conducted, to extract information on the transient rate
for every combination of flux and duration of the transients.
Our simulations can be extended to take into account realistic scenarios, like multiple pointings and variable noise, by assigning
spatial coordinates to the simulated transients.

\section*{Acknowledgments}
The authors would like to thank the referee for the useful comments that helped to improve the paper.
DC and RAMJW acknowledge support from the European Research Council Advanced Grant
247295 ``AARTFAAC'' (P.I. R. Wijers).

\bibliographystyle{mnras.bst}
\bibliography{/Users/dariocarbone/Documents/carbone_bibliography.bib}

\end{document}